\documentclass[a4paper, biblatex, keeplastbox, pdftex]{jacow}

\usepackage[USenglish]{babel}

\newcommand{\Ni}{(1)~}
\newcommand{\Nii}{(2)~}
\newcommand{\Niii}{(3)~}

\begin{document}
\title{FastWARC: Optimizing Large-Scale Web Archive Analytics}

\author{Janek Bevendorff$^*$, Martin Potthast$^\dagger$, Benno Stein$^*$\\
$^*$Bauhaus-Universit{\"a}t Weimar, $^\dagger$Leipzig University}

\maketitle

\begin{abstract}
Web search and other large-scale web data analytics rely on processing archives of web pages stored in a standardized and efficient format. Since its introduction in~2008, the IIPC's Web ARCive (WARC) format%
\footnote{ISO\,28500:2017; \url{https://iipc.github.io/warc-specifications/}}
has become the standard format for this purpose. As a list of individually compressed records of HTTP requests and responses, it allows for constant-time random access to all kinds of web data via off-the-shelf open source parsers in many programming languages, such as \mbox{WARCIO},%
\footnote{\url{https://github.com/webrecorder/warcio}}
the de-facto standard for Python. When processing web archives at the terabyte or petabyte scale, however, even small inefficiencies in these tools add up quickly, resulting in hours, days, or even weeks of wasted compute time. Reviewing the basic components of WARCIO and analyzing its bottlenecks, we proceed to build {\it FastWARC}, a new high-performance WARC processing library for Python, written in C++\,/\,Cython, which yields performance improvements by factors of~1.6--8x.
\end{abstract}

\vspace{-0.5ex}
\section{Introduction}

The earliest open source implementations of the WARC format were provided for Java, namely Lin's {\it ClueWeb Tools}%
\footnote{\url{https://github.com/lintool/clueweb}}
(initially used by the research search engine ChatNoir),%
\footnote{\url{https://chatnoir.eu/}}
followed by a more standards-compliant reference implementation from the IIPC.%
\footnote{\url{https://github.com/iipc/jwarc}}
Meanwhile, the IR, NLP, and machine learning communities have largely transitioned to Python, instead adopting WARCIO as a native implementation in that language. Processing large samples of the Common Crawl and web archive data from the Internet Archive, however, we observed that the library did not match our performance expectations. Even compiling it to native C~code using Cython yielded only marginal improvements. Analyzing its bottlenecks, three key causes can be discerned:
\Ni
stream decompression speed,
\Nii
record parsing performance, and
\Niii
lack of efficient skipping of non-response records.
Our contribution is to rectify these issues with {\it FastWARC}, a rewrite of the entire WARC parsing pipeline from scratch.

\vspace{-0.5ex}
\section{FastWARC vs. WARCIO}

FastWARC is a reimplementation of \mbox{WARCIO} in~C++ with Cython, making it both fast and perfectly integrated into the Python ecosystem, yet allowing for more language bindings if required. Table~\ref{table-performance} compiles detailed performance comparisons: On an uncompressed WARC file, it gains an overall 6.4x~speedup over \mbox{WARCIO}, or 4x over a naively ``cythonized'' \mbox{WARCIO}. With an average processing time of 1.2~vs.\ 8~seconds for a single WARC file, this already saves at least 115~hours of compute time on a recent Common Crawl with 64\,000 individual WARCs (62.5\,TiB compressed). For better decompression speed of gzipped streams, FastWARC interfaces directly with {\it zlib}, achieving compute time savings of roughly 2.1~hours per~TiB or 2\,200~hours per~PiB over \mbox{WARCIO}. The largest performance penalty, however, comes from the decompressor itself. While still saving about 135~hours overall on a Common Crawl, the relative speedup shrinks to only~1.6--1.8x. For this reason, we decided to add support for the more recent and much faster LZ4 algorithm. With LZ4, we can save another 168~hours on top (a speedup of 4.8x over FastWARC with GZip), or 300~hours compared to \mbox{WARCIO}~(speedup of up to~8x).
\begin{table}
\setlength{\tabcolsep}{9.9pt}%
\footnotesize%
\begin{tabular}{@{}ll@{\hskip 10pt}rr@{}}
\toprule
\bf Comp.\ & \bf Parser & \bf Records/s & \bf Speedup \\
\midrule
\multicolumn{4}{@{}l@{}}{\it AMD Ryzen Threadripper 2920X (NVMe SSD)}\\
\midrule
None   & WARCIO                 &  16\,945.5  & -- \\
None   & FastWARC               & 108\,488.0  & 6.4 \\
None   & WARCIO+HTTP            &  11\,661.6  & -- \\
None   & FastWARC+HTTP          &  79\,297.0  & 6.8 \\
None   & WARCIO+HTTP+Checksum   &   6\,986.7  & -- \\
None   & FastWARC+HTTP+Checksum &  21\,320.9  & 3.1 \\
\midrule
GZip   & WARCIO                 &   6\,460.1  & -- \\
GZip   & FastWARC               &  10\,413.4  & 1.6 \\
GZip   & WARCIO+HTTP            &   5\,435.6  & -- \\
GZip   & FastWARC+HTTP          &  10\,101.5  & 1.9 \\
GZip   & WARCIO+HTTP+Checksum   &   4\,121.6  & -- \\
GZip   & FastWARC+HTTP+Checksum &   7\,433.0  & 1.8 \\
\midrule
LZ4   & FastWARC                &  49\,825.4  & 7.7$^*$ \\
LZ4   & FastWARC+HTTP           &  42\,394.5  & 7.8$^*$ \\
LZ4   & FastWARC+HTTP+Checksum  &  16\,992.2  & 4.1$^*$ \\
\midrule
\multicolumn{4}{@{}l@{}}{\it Intel(R) Xeon(R) CPU E5-2620 v2 (remote Ceph storage)}\\
\midrule
None   & WARCIO                 &   7\,969.1  & -- \\
None   & FastWARC               &  49\,396.5  & 6.2 \\
\midrule
GZip   & WARCIO                 &   3\,555.7  & -- \\
GZip   & FastWARC               &   6\,335.1  & 1.8 \\
\midrule
LZ4   & FastWARC                &  28\,313.8  & 8.0$^*$ \\
\bottomrule
\end{tabular}
\vspace{.5ex}
\caption{Evaluation of FastWARC and WARCIO on two systems. Runs are
\Ni
without payload parsing,
\Nii
with automatic HTTP header parsing, and
\Niii
with record checksumming.
$^*$LZ4 speedup is over WARCIO with GZip, since WARCIO does not support LZ4.}%
\label{table-performance}%
\vspace{-7ex}
\end{table}

\vspace{-.4ex}
\section{Conclusion}

FastWARC can speed up WARC processing significantly, saving hundreds of hours of compute time on large-scale web archive analytics. By far the largest speedup, though, can be gained from using LZ4 over GZip. Considering an additional storage overhead of only about 30--40\,\%, recompressing GZip WARCs with LZ4 is certainly an option to be considered, especially in cases where processing speed is more important than storage efficiency.

FastWARC is released under the Apache~2.0 license and can be downloaded from Github%
\footnote{\url{https://github.com/chatnoir-eu/chatnoir-resiliparse}}
or PyPi.%
\footnote{\tt\footnotesize pip install fastwarc}
\vspace{-.2ex}

\end{document}